\documentclass{article}

\usepackage{graphics}
\usepackage{graphicx}

\newcommand{\AS}{$AdS_5 \times S^5${}}
\newcommand{\N}{$\mathcal{N}$}
\newcommand{\ra}{\rightarrow}
\newcommand{\lth}{\lambda_{\mathrm tH}}
\newcommand{\f}{\begin{equation}}
\newcommand{\ff}{\end{equation}}
\begin{document}

\title{The Maldacena conjecture and Rehren duality}

\author{Matthias Arnsdorf$^1$\thanks{arnsdorf@nbi.dk} {}\ and Lee
Smolin$^{2,3}$\thanks{smolin@phys.psu.edu} \\
\smallskip\\
{\small ${}^1$ The Niels Bohr Institute, Blegdamsvej 17, DK-2100 Copenhagen \O.} \\
{\small ${}^2$ Center for Gravitational Physics and Geometry}\\
{\small  Department of Physics, The Pennsylvania State University,}\\
{\small University Park, PA, USA 16802.}\\
{\small ${}^3$ The Blackett Laboratory,} 
{\small Imperial College of Science, Technology and 
Medicine}\\
{\small London SW7 2BZ, UK} 
}

\date{June 6, 2001}

\maketitle

\begin{abstract}

We analyse the implications of the fact that there are two claims for
a dual to \N=4 superconformal SU(N) Yang-Mills theory (SCYM), the
Maldacena conjecture and the theorem of Rehren. While the Maldacena
dual is conjectured to be a non-perturbative string theory for large
string coupling $g_s$ and small $N$, the Rehren dual is an ordinary
quantum field theory on $AdS_5$ for all values of the parameters. We
argue that as a result, if we accept the Maldacena conjecture, one of
the following statements must be true: 1) SCYM does not satisfy the
axioms of algebraic quantum field theory for finite $N$ because its
observables do not obey the causal structure of conformal Minkowski
space;.  2) String theory on \AS\  is not a quantum theory of gravity in
$10$ dimensions because it is dual to an ordinary quantum field theory
on $AdS_5$ whose causal structure remains fixed for all values of its
couplings; or 3) there is no consistent quantisation of string theory
on $AdS_5 \times S^5$ for finite string scale $l_s$ and $g_s$.

In evaluating the evidence for each of these conclusions 
we point out that many of the
tests of the Maldacena conjecture can be explained by a weaker
form of an $AdS/CFT$ correspondence.

\end{abstract}

\section{Introduction}

In 1997 Maldacena~\cite{juan} made a bold
conjecture that string theory on \AS\  is actually equivalent under
a certain duality relation to a quantum field theory on
the four dimensional Minkowski spacetime which is a
time-like boundary for $AdS_5$.  This was quickly reinforced
by arguments coming from a number of directions~\cite{adscft1,adscft2},
among them the
fact that it seemed to offer striking confirmation of the
holographic principle.  It was seen
also to be closely related to older ideas of Polyakov
on the role of  additional dimensions in conformally invariant
theories as well as old results from studies of supergravity on
$AdS_5$~\cite{adscftreview}.

This led to a very large number of papers,
many of which do confirm the existence of a correspondence between
the predictions of \N=4 Super-Yang-Mills theory and linearised
supergravity on a background which is \AS. Some
of the checks are quite striking in their precision. There are also
very interesting results that show that Yang-Mills theory with fewer
supersymmetries are related by the same kind of correspondence to
other supergravity theories, or to linearised supergravity on different
backgrounds.

Among the papers inspired by the Maldacena conjecture are a
series by Rehren~\cite{Rehren:2000jn,Rehren:2000tp,Rehren:1999um}
 proving a theorem  which states
 that any algebraic quantum field theory on an $AdS$ space of any dimension
 is dual to an explicitly constructible, conformal, algebraic field theory on
 the boundary. This appears
to be highly relevant to Maldacena's conjecture. In the limit in
which gravity is weak, Rehren's duality seems to confirm the
predictions of the Maldacena conjecture, and may even be said to
provide a simple explanation for some of the results which
support the Maldacena conjecture.

It would seem remarkable if there were two independent dualities
between field theories on $AdS$ and its boundary.
However, the exact relationship between the theorem and the
conjecture is not completely clear. The issue, which it is the
purpose of this note to explore, is that the duality proved by
Rehren connects the \N=4 supersymmetric
Yang Mills theory to an ordinary local quantum field theory on
the fixed $AdS_5$ background {\it for all values of coupling and
for all $SU(N)$.}  This is puzzling as the Maldacena conjecture
appears to require that the bulk theory should be a string theory
when $g_s \approx g_{YM}^2$ is large and $N$ is small. But if we
use both dualities we are able to predict that this string theory
is dual to an ordinary quantum field theory on the fixed $AdS_5$
background, as both are dual to the supersymmetric Yang-Mills
theory. If true, this is  surprising, as it seems to imply
that a string theory is dual to an ordinary quantum field theory
in a space of the same dimension, not one lower. It also seems to
imply that the causal structure does not become dynamical once
gravity and quantum gravitational effects become important.

In this note we offer what seems to us to be the most
logical reading of the possible implications of the
existence of both Rehren's theorem
and the conjecture of Maldacena. We find that present knowledge
does not allow us to distinguish between several
alternatives, each of which has non-trivial consequences.

\section{The Maldacena conjectures}

The Maldacena conjecture~\cite{juan,adscft1,adscftreview}
identifies type IIB string theory on \AS\  with \N = 4
superconformal-Yang-Mills theory with gauge group $SU(N)$ (from
now on SCYM). We will distinguish between  weak and
 strong versions of the conjecture. First we list the relevant
constants of the theory and their proposed relations:

\bigskip
\begin{tabular}{llll}
$g_s$ & string coupling constant & $g_{YM}$ & Yang-Mills coupling constant \\
$l_s$ & string scale & $N$ & dimension of gauge group SU(N) \\
$G_N$ & Newton's constant & $R$ & radius of curvature of \AS
\end{tabular}

\bigskip
This leads to the following relations:
\begin{eqnarray*}
g_{YM}^2 &\sim& g_s\\
\lth = g_{YM}^2 N \sim g_s N &\sim& R^4 / l_s^4,
\end{eqnarray*}
where $\lth$ denotes the 't Hooft coupling.

It helps to divide the Maldacena conjecture into several forms,  from
weaker to stronger.
The weak form of the Maldacena conjecture can be stated as:

\begin{description}
\item[M1:] In the limit $N \ra \infty$, $g_{s}$ small,
$ \lth$ large, SCYM on
the 4-d Minkowski-space boundary $M_4$ of \AS\  becomes equivalent to
type IIB supergravity on \AS, defined by a power series expansion
with fixed metric on \AS.
\end{description}

Note that this is necessarily a duality with  supergravity, {\it defined
in terms of a power series expansion around the fixed \AS\ 
metric.} To any finite order this can be considered
a field theory on \AS\  since the limit of very small $g_s$ implies
that string theory may be neglected.
We may note that this weak form of the Maldacena conjecture is
confirmed by many calculations, there is also no conflict with
Rehren's results, as we will describe below.  We must also emphasize 
that the relationship is not necessarily a duality, at this level 
there is no conjecture that all of type IIB supergravity is contained 
in SCYM.  

The next strongest form of the Maldacena conjecture concerns
the case where we consider the full nonlinearities
in classical gravitational effects, while still
suppressing quantum gravity and string effects.
We do this by considering arbitrary $\lth$ while
still keeping $g_s$ very small. We then have what we may call the
``medium" Maldacena conjecture:

\begin{description}
\item[M1.5:] SCYM with arbitrary $N$ but $g_{YM}$ very small must be
equivalent to classical supergravity on
spacetimes which are \emph{asymptotically} \AS.
\end{description}

The strong form comes from considering arbitrary values
of the parameters $N$ and $g_{YM}$:

\begin{description}
\item[M2:] For all values of
$N$ and $g_{YM}$, SCYM on $M_4$ is dual to type IIB string
theory on spacetimes which are \emph{asymptotically} \AS.
\end{description}

In the medium and strong forms of the conjecture we can only require
the spacetime on which the string theory lives to be asymptotically
\AS\ since we are no longer studying weak gravitational and matter
degrees of freedom propagating on a fixed causal structure given by
the metric on \AS. This
is necessary if the theory is to describe structures like black holes
where the deviations from a fixed \AS\ are arbitrarily
strong.

Indeed, the idea behind the
strong Maldacena conjecture is that the Yang-Mills theory
``effectively sums over all spacetimes'' which are asymptotic to \AS,
thus providing a non-perturbative definition of string theory.
Furthermore, different pure and statistical states of SCYM are believed
to correspond, in the classical limit, to different asymptotically \AS
spacetimes.  For example time dependent metrics, such as those with
gravitational waves are believed to correspond to non-equilibrium or
time dependent states of the SCYM theory.

Hence,  any theory which has general relativity or supergravity
as a classical limit ($\hbar \ra 0, G_N$ fixed) --- which is a
minimal requirement for a quantum theory of gravity --- cannot
have a fixed causal structure, otherwise we come into conflict
with basic predictions of general relativity (now well confirmed),
such as light bending, gravitational lensing etc.

In summary, we require:
\begin{description}
\item[BI:]  A quantum theory of gravity must be {\it background
independent} so that its classical limit is general relativity or a
generalisation thereof, whose metric and hence causal structure
are dynamical.  Such a theory may be constrained by boundary conditions
that depend on a metric asymptotically,
in the same sense that general relativity can be formulated in terms
of asymptotically flat boundary conditions.  But the causal structure
in the bulk cannot be fixed, and must be different with different
solutions or quantum histories of the spacetime.
\end{description}

Thus, we should make explicit the following, which is usually
assumed in the statement of the Maldacena conjecture:

\begin{description}
\item[M3:] Type IIB string
theory on spacetimes which are \emph{asymptotically} \AS\ for all values of
$l_s$ and $g_s$ is a quantum theory of gravity, in the sense that
it has the full solution space of supergravity with \emph{asymptotically} \AS
as a classical limit and hence it satisfies the property {\bf BI}.
\end{description}

\section{Algebraic field theory and AdS/CFT} \label{RD}

Rehren's papers~\cite{Rehren:2000jn,Rehren:2000tp,Rehren:1999um}
assume that SCYM is an algebraic quantum field theory (AQFT).
This is the most general structure for a quantum field theory
that incorporates relativistic covariance and local causality
(c.f.~\cite{Buchholz:2000xm} for a review and references).
The basic objects in AQFT are the localised algebras of
observables. More specifically an AQFT is completely specified by
a assignment of an algebra $A(O)$ to every open region $O$ of
spacetime. Furthermore these algebras have to comply with
covariance, locality and isotony as is captured in the following
axioms:
\begin{enumerate}
\item  The algebras localised at  spacetime regions related by  a
space-time symmetry transformation $g$ (in our case this will be
$SO(2,4)$) are unitarily equivalent i.e.: 
$A(gO) =U(g)A(O)U(g^{-1})$, where 
$U(g)$ is a representation of the symmetry group.

\item Two observables localised at space-like distances commute.

\item An observable localised in a region $O$ is localised in a larger
region $O' \supset O$, i.e.: 
$O' \supset O \Rightarrow A(O') \supset A(O)$.

\end{enumerate}

It has been shown that this local algebra structure suffice to
recover standard quantum field theory. In particular states arise as
the carriers of a chosen representation of the net of algebras. Point
fields operators $\phi(x)$ can be recovered, roughly, by taking the
intersections of the algebras localised around $x$:
\[
\{\phi(x)\} = \cap_{x \in O} A(O),
\]
where the above has to be understood in an appropriate technical sense
to be meaningful.
The $AdS/CFT$ correspondence relies on the fact that the isometry group
of $AdS_{1,d}$ in d+1 dimensions and the conformal group of Minkowski
space-time are both $SO(2,d)$. Furthermore the boundary of $AdS_{1,d}$ is
d-dimensional conformal Minkowski space $M_{d}$ and the restriction of
the $AdS$ group to the boundary acts like the d-dimensional Minkowski
conformal group.

Remarkably, Rehren~\cite{Rehren:2000jn,Rehren:2000tp,Rehren:1999um}
has been able to show that this symmetry property is in fact
sufficient to establish a duality between any conformal AQFT on $M_d$
and an AQFT on $AdS_{1,d}$, and also, conversely, a duality between
any AQFT on $AdS_{1,d}$ and a conformal AQFT on the boundary.  Note
that this duality is not a conjecture but a rigorous theorem of
algebraic quantum field theory.  The duality is unique and provides an
explicit map between observables of the two theories.  Furthermore,
the duality does not refer to string theory or quantum gravity and
hence is not related to standard versions of the holographic
principle.

The idea is to map observable algebras on
$M_d$ to algebras on $AdS_{1,d}$ in such a way that causality,
covariance and isotony are preserved. This gives us a unique net of
algebras on $AdS_{1,d}$, which in turn defines a quantum field theory
with the causal structure of $AdS$. Since the observable algebras for
the theory on $M_d$ and the theory on $AdS_{1,d}$ are the same it
follows that they will have the same representations and thus that the
state spaces of the two theories will coincide - hence the theories
are dual. But crucially the space-time localisation of the algebras is
very different in the two theories. This leads to different
interpretations of the observables. In particular, the Hamiltonian on
$AdS_{1,d}$, given by the generator of time translations, is identified
with the linear combination  of translations and conformal
transformations  in the 0-direction of $M_d$: $\frac{1}{2}(P^0 +K^0)$.
This allows for the mathching of the degrees of freedom 
in the bulk and the boundary theory.

In more detail, the correspondence is given as follows. Elementary
regions in $M_d$ are double cones $K$, i.e.\ the intersections of a future
and past-directed light cone. These are causally complete convex
regions. Each double cone uniquely determines a wedge region $W$ in
$AdS_{1,d}$ which is the causal completion of $K$ in $AdS$, i.e. all
points from which one can receive signals from a point in $K$ and send
signals to a different point in $K$. Given a net of algebras on $M_d$
this allows us to construct a net on $AdS_{1,d}$ via: $A(W) = A(K)$ 
if $K$ corresponds to $W$. The key to Rehren's result is to show that
this correspondence preserves causality, covariance and isotony. Given
the algebras defined on the wedges $W$ we can now associate an algebra
to any region $O$ in $AdS_{1,d}$ via:
\[
A(O) = \cap_{O \subset W} A(W).
\]
This defines a unique AQFT on $AdS_{1,d}$ which is dual to the
original AQFT on $M_{d}$.

It is important to note that the above duality is between local
algebras. Point-like field operators, in general, are not mapped to
other field operators.
Indeed, an important property of the above map is that genuine field
operators on $M_d$ are dual to observables in $AdS_{1,d}$ that are
attached to the boundary. Bulk localised observables in $AdS_{1,d}$ on
the other hand are mapped to `extended' observables on the
boundary. Such observables cannot be described in terms of fields and
could, for example, correspond to Wilson loops in a boundary Yang-Mills
theory.

Schroer~\cite{Schroer:1999tv} has used this property of the correspondence
to suggest that the
locality structure of a field theory on $AdS$ dual to a SCYM theory on
the boundary is not compatible with that of string theory or
(linearised) supergravity.  We will not investigate this further,
instead we focus on the fact that any SCYM theory on $M_4$ is dual,
via Rehren's map, to
a theory with the $\emph{fixed}$ causal structure of $AdS_5$.

\section{Consequences}

SCYM is a conformal quantum field theory on $3+1$ dimensional Minkowski space.
In this section we assume that the observables of SCYM obey the casual
structure of $M_4$ and hence satisfy the axioms of algebraic field
theory. Let us make this explicit,
\begin{description}
\item[AQFT:]  ${\cal N}=4$ supersymmetric Yang Mills theory satisfies
the axioms of algebraic quantum field theory for all $SU(N)$ and
all values of the gauge coupling.
\end{description}
 In this section we draw out the implications of the above in
conjunction with the various versions of the Maldacena conjecture.

Rehren duality  allows us to construct uniquely an algebraic quantum
field theory (the \emph{Rehren dual} in the following) on $4+1$ dimensional
AdS space which is dual to SCYM on its $3+1$ dimensional
Minkowski space boundary.
This duality is valid  for any values of the parameters $N$
and $g_{YM}$. Furthermore, by definition, the AQFT on $AdS$ is
defined with respect to a fixed causal structure which is reflected in
the commutation relations between any two observables. Hence the
theory will not satisfy property {\bf BI}.
 This leads to the following  conclusion:

\begin{description}
\item[Conclusion 1:]
Any theory that is dual to
SU(N) superconformal Yang-Mills theory on $M_4$
is also dual to an ordinary field theory defined on the fixed $AdS_5$ background.
\end{description}

\begin{figure}
\begin{center}
\includegraphics*[height=3cm]{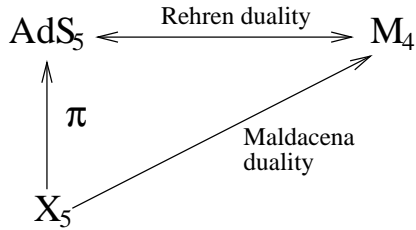}
\caption{Composition of dualities to SCYM theory.} \label{duality}
\end{center}
\end{figure}

The Maldacena conjectures relates the SCYM on $M_4$ to a theory
on a manifold $X_5$ which is asymptotically \AS. The above
implies that this theory is dual --- via a duality $\pi$ = (Rehren
duality)$\circ$(Maldacena duality) --- to a field theory with the
fixed \emph{global} causal structure of $AdS_5$. This is
illustrated in figure~\ref{duality}.
We now examine the implications for each version of the Maldacena conjecture.

\begin{itemize}

\item[{\bf M1:}] If we assume conjecture {\bf M1} then there is no
contradiction since linearised supergravity in the $g_s \ra 0$ limit
is a field theory with a fixed background. In this case the Maldacena
duality and the Rehren duality could coincide.  Essentially the
representation theory of the supersymmetric extension of
SO(4,2) includes both the states of SCYM and the linearised
modes of supergravity on \AS.

\item[{\bf M1.5:}] Conjecture {\bf M1.5} can be refuted since it implies that
the space of asymptotically $AdS$ solutions to classical general relativity 
can be described equivalently by an ordinary  field theory on $AdS$ with a fixed
metric and causal structure.  But this is contradicted by the  fact that
in general relativity the causal structure varies dynamically, which
has been observed in many light bending experiments.

\item[{\bf M2:}]  If we assume conjecture {\bf M2} then type IIB
string theory cannot satisfy property {\bf BI}. Therefore {\bf M3}
fails and string theory
cannot be a theory of quantum gravity.
\end{itemize}

It is worth mentioning that these conclusions need
not be in contradiction with the so called
``uniqueness" theorems that show, subject
to various assumptions, that any theory whose linearisation is
general relativity has the interactions given by general
relativity or some extension of it. First,
we should note that this theorem has
been proven only under restricted
assumptions
\cite{Deser:1970wk} and  explicit counterexamples are known in
$3$ and $5$ dimensions \cite{Boulanger:2000ni}.  So it is possible
that these theorems do not apply in the special case under
consideration.

However, even if this is not the case, it must be emphasised
that the existing uniqueness theorems only establish the
forms of the interactions and gauge symmetry order by order
in a power series expansion of the classical theory on a fixed
background.  Whether such a power series expansion
is equivalent to the full non-linear gravitational theory is
an additional issue, not addressed by the existing theorems.  It
should further be stressed that the expansions in question are
generally asymptotic rather than converging power series.
Thus, it is possible that {\bf M1} may hold to
some finite order of an expansion in small disturbances around
the fixed \AS\ background, but that still
{\bf M1.5} and {\bf M2} fail. In this case the Rehren and
Maldacena dualities could agree to any finite order, but
SCYM would only give a construction of a classical gravitational theory
good to a finite order in an expansion in small amplitude
excitations propagating on the fixed \AS\ background.
This would agree with many of the checks which have been performed,
but there is then no necessity that {\bf M2} holds or that the Maldacena
duality gives a {\it non-perturbative} or
background independent definition of string theory
in terms of the SCYM theory.

There is also the possibility that {\bf M2} could fail  because
string theory on \AS\ fails to exist as a quantum theory for all
values of $l_s$ and $g_s$. Despite the large literature
on the Maldacena conjecture it has proved so far surprisingly
difficult to construct the quantum string theory on \AS, even in
the case of the free string.  The present situation~\cite{arkady}
is that classical actions for string theory on \AS\ have been
constructed and it has been shown that their 
$l_s \rightarrow 0$ limits are described by the superparticle on \AS.
Hence, one can deduce that the quantisation of the 
$l_s \rightarrow 0$ limit reproduces the spectrum of linearised
supergravity on \AS.  However for finite $l_s$ the
worldsheet action is non-linear and there has been as yet no
successful quantisation of the theory, even at the level of the
free string.  Hence the spectrum of the free string is not known
and it is not known for certain that there actually is a
consistent string theory on \AS.

To summarise the logic, we reach the conclusion either that {\bf AQFT} fails
or at least one of the conclusions reached in this
section must hold. In particular, if {\bf AQFT} holds then 
{\bf M1.5} and {\bf M2} or {\bf M3} must fail.

\section{Conformal Induction}

Given that a large number of results have been offered in support of
one or another version of the Maldacena conjecture, the reader may be
tempted to conclude from the foregoing that {\bf AQFT} fails. However
before reaching that conclusion we should check whether the evidence
adduced in favour of a version of the Maldacena conjecture can be
explained by a weaker hypothesis.  This is what we discuss in this
section. We may note that the arguments that follow provide one
\emph{possible} resolution of the conflict between Rehren duality and
the Maldacena conjecture. The arguments of the previous section
hold regardless of whether we chose to adopt this explanation.

Witten~\cite{adscft1} and Gubser, Klebanov and Polyakov \cite{adscft2}
have offered a correspondence between a quantum
theory on a $d+1$ dimensional spacetime $X_{d+1}$, which is
asymptotically anti-DeSitter and a conformal field theory on the
spacetime $M_d$ which is the conformal boundary of $X_{d+1}$. This
version of the $AdS/CFT$ correspondence is considerably weaker and
more general than the conjecture of Maldacena.

Let us review the prescription, which allows us to construct the
correlation functions of a conformal field theory on $M_d$  by
evaluating the
 correlation functions of fields of the $d+1$ dimensional theory
 in the limit in which the fields are taken to the boundary.

More precisely, let 
  $\phi$ be a field of the $d+1$ dimensional theory and
${\cal O}$ an operator of the $d$ dimensional conformal field
theory, which is chosen to correspond to it. Then the
$CFT$ is defined by the following formula~\cite{adscft1}
\f
\langle e^{\int_{S^d}\phi_0 {\cal O}}\rangle_{CFT} = Z (\phi )
\label{CI}
 \ff
 where $Z (\phi)$ is the partition 
 function on  $X_{d+1}$ with $\phi$ constrained
 to approach fixed values $\phi_0$ on the boundary
 $M_d$.  
 
 We say the boundary theory is induced by the bulk theory through its
 definition in terms of correlation functions.  We call the boundary
 theory the {\it conformal image} of the bulk theory.
 
As already mentioned conformal induction is a general procedure that
requires only that the bulk theory be a well defined (Euclidean or
Lorentzian) quantum theory on a fixed manifold.  There is no necessary
relationship with gravity, supersymmetry\footnote{ Supersymmetry may
be necessary in some cases to guarantee that the bulk partition
function exists, but it need not be necessary in all cases.  There are
indeed low dimensional examples of conformal induction in which
supersymmetry plays no role such as in the case of $2+1$ gravity.} or
string theory.
If  the partition function $Z(\phi )$ exists and the fields $\phi$  have well
defined boundary values then the symmetries of the asymptotically $AdS$
manifold are sufficient to show that the field theory defined by
(\ref{CI}) will satisfy the symmetries of a conformal\footnote{The induced
theory is conformal because the asymptotically $AdS$ metric defines the
boundary metric only up to a conformal factor.}  quantum field theory.

We note however, that in the case that the bulk field theory is defined
on a \emph{Lorentzian} manifold and $M_d$ is a Minkowski spacetime,
there may be an obstruction to conformal induction.  Additional
consistency conditions  must be satisfied if the induced theory
is to satisfy unitarity and locality.  This follows from the
possibility that events on the boundary which are spacelike with
respect to the metric on the boundary may be causally connected
through causal geodesics in the bulk.  This can occur, for example if
there are closed time like curves in the bulk.  Some examples of this
phenomena relevant for supersymmetric theories are described in
\cite{Caldarelli:2001iq}. Hence we need the {\it causal consistency condition}
(CCC), which states that {\it two events are causally related in
the boundary Minkowski spacetime if and only if they are causally
related in the metric of the bulk theory.}
     
We come now to the crucial point, which is that there is no reason
that the above procedure, which is very general, should provide us
with a duality between theories. In particular, conformal induction is
a one way process. There is no known general procedure for
reconstructing the bulk theory from the induced boundary theory, nor
should the relationship between the bulk and the boundary theories be
unique.  Conformal induction involves only the boundary data of the
bulk quantum theory. There is no requirement that all observable
quantities of the bulk are computable in the conformal 
image. This
applies especially to observables which are completely localised in
the bulk\footnote{It is clear from the Penrose diagram of $AdS$ that data 
on the time like boundary is not sufficient to determine a classical 
soloution to the field equations, data must be given as well at past 
infinity.}.

In the case of a Lorentzian bulk manifold $AdS_{d+1}$ we can ask about
the relation between conformal induction and Rehren duality. These
need not necessarily be related, however this would result in there
being two different boundary conformal field theories corresponding to a
given theory on $AdS_{d+1}$. More interesting is the possibility that
the conformal image and the Rehren dual coincide. In this case the
conformal induction would provide a true duality with an inverse. But
as described at the end of section~\ref{RD} the Rehren dual maps bulk localised
observables to extended objects in the boundary. It is unclear how
this can be achieved via the conformal induction procedure. More
likely is the possibility~\cite{Rehrenprivate} that the conformal
image is precisely the sub-theory of the Rehren dual that contains the
true field observables, which according to Rehren duality are related
to bulk observables attached to the boundary.
 
\subsection{Conformal Induction and the Maldacena conjecture}

We now turn to the implications of the above for the Maldacena
conjecture. This is examined most easily by formulating the conformal
induction conjecture:
\begin{description}
\item[CI:] The correlation functions of observables in SCYM theory
on the Minkowski space boundary $M_4$ can be evaluated via
equation~(\ref{CI}) using the partition function of the type IIB
string theory on the bulk manifold, which is asymptotically \AS.
\end{description}
As with the Maldacena conjecture this can also be formulated in
weaker versions in terms of linearised supergravity theory on \AS,
in which it holds only in the limits of $N \ra \infty$, $g_{s}$ small,
$ \lth$ large.

The essential point is that, as explained above, {\bf CI} is weaker than
any of the versions of 
the Maldacena conjecture, in particular it does not posit a duality.
For example, {\bf M2} is equivalent to the conjecture {\bf CI}
together with the claim that there is a unique inverse to the
conformal induction procedure\footnote{ We may note that in the
Lorentzian case the CCC must be imposed as a restriction on the sum
over manifolds, otherwise the induced theory will violate causality on
the boundary Minkowski spacetime.}, so that all observable quantities
in IIB string theory can be computed from the SCYM theory.

This implies, however, that any tests of the relation between SCYM and
IIB string theory
that can be explained by the hypothesis {\bf CI} are not
sufficient to provide support of the much stronger Maldacena
conjecture. Let us examine some of the evidence that has been adduced:

\begin{itemize}
    
    \item The matching of $N$ point functions between
    classical scattering in linearised supergravity on
    \AS\ and correlation functions of
    SCYM.  This is clearly explained by {\bf CI}
     to any finite order in a power series expansion in $g_s$ around
\AS. Note that in this case we also have the possibility that the
conformal image is the Rehren dual or a subset of it.

    \item Various results regarding the scaling behaviour of
    field theories in four dimensions in terms of the behaviour
    of classical supergravity in asymptotically $AdS$ spacetimes.
    These are very useful but again are explained by conformal
    induction and so offer no independent support for an equivalence
    of the kind postulated in the Maldacena conjecture. 
    
    \item Various results in $2+1$ dimensions where there are no local
    degrees of freedom.  Given the special nature of gravitational
    physics in $2+1$ dimensions these illuminate but cannot be taken
    as strong evidence for conjectures about higher dimensional
    theories.
    
     \item The matching~\cite{adscft1} (up to an overall constant) of
    the entropy of large black holes in asymptotically $AdS_5$
    spacetimes with thermal states of SCYM. Again this only provides
    evidence for {\bf CI}. In this case the boundary manifold is $S^1
    \times S^3$ so that the induced boundary theory is a thermal
    quantum field theory defined at the temperature given by the
    inverse period of the $S^1$.  

\item However, in addition the
    entropies of two different phases of SCYM have been argued to
    match the entropies of field theories on different asymptotically
    $S^1 \times S^3$ manifolds~\cite{adscft1}.  This suggests that, in
    the Euclidean case, fluctuations around different bulk manifolds
    are necessary to match supergravity to all the physics of
    SCYM. This appears at present to be the strongest evidence for
    {\bf M1.5} and {\bf M2}, or at least for a form of {\bf CI} in
    which the partition function of a full quantum theory of gravity
    is required to induce the full physics of SCYM.  To the extent
    that the only known candidate for  such a theory whose
    semiclassical limit reproduces supergravity is string theory, this
     provides evidence for a correspondence between string theory
    and SCYM at the level of {\bf CI}.

\end{itemize}

This is an incomplete list, but it shows that care must be taken in
evaluating the significance of various results offered as evidence
for the Maldacena conjecture.

\section{Summary}

We have seen that the Maldacena conjecture in conjunction with Rehren's
theorem leads to a duality $\pi$ between string theory and a quantum
field theory on a fixed $AdS_5$ background. This is consistent in the
linearised supergravity limit and hence with conjecture {\bf M1} but
 there
appears to be a problem once $g_s$ and/or $\lth^{-1}$ are allowed
to take on arbitrary values. The arguments sketched above have led to
the conclusion that one of the following must be true:

\begin{enumerate}

    \item{} {\bf AQFT} fails  for finite $N$
    and $g_{YM}$.  In this case Rehren's theorem is not relevant and there
    is no contradiction with the strong Maldacena conjecture.
     However,  the axioms of AQFT are very general. Unless SCYM
fails altogether to be a good quantum theory, it seems the only
way it  could fail to be an AQFT is if it violates causality on
the four dimensional Minkowski space.

    \item{} {\bf M2} fails, possibly because there is no interacting quantum
    string theory on \AS\ for finite $l_s$.

    \item{} {\bf M3} fails. In this case quantum string theory exists
    on \AS\ for finite $g_s$ but it is not a quantum theory of gravity,
    because property {\bf BI} fails. This comes about because the
    combination of {\bf M2} and the Rehren theorem implies it can be
    expressed in terms of a dual description based on an AQFT on the
    fixed metric and causal structure of \AS.

\end{enumerate}

Further, we noted that many of the tests carried out on the $AdS/CFT$
correspondence
can be explained by the hypothesis {\bf CI}. Since {\bf CI} is 
logically weaker and more general than any form of the Maldecena
conjecture, these tests do not provide independent evidence for 
any part of the Maldecena conjectures that goes beyond {\bf CI}.  In 
particular, this leaves the possibility that SCYM captures some 
information about certain observables in supergravity, without there
being a duality or equivalence of the theories. 

This leaves us with the question of the status of {\bf M1}, 
as it is the one form of the Maldecena
conjecture that does not run into conflict with the conjunction of
{\bf AQFT} and Rehren's theorem. 
But as {\bf M1} involves only a theory
on the fixed background \AS\ it is then possible to ask whether 
it may be in fact a consequence of
Rehren's theorem, i.e.\ whether the conformal image of linearised
supergravity is given by the Rehren dual or a subset of it.
 This seems plausible since it would 
be highly
remarkable if there were two independent dualities between SCYM
and a field theory on $AdS$. If this were the case the connection
to a gravitational theory might be in some sense accidental.

Indeed, while Rehren's theorem demonstrates explicitly that two theories
living on spaces of different dimensions can be dual, it also implies that
this duality need not involve gravity or holography.
In the present case the duality
would be a consequence only of the fact that the supersymmetric
extension of $SO(4,2)$ has representations that include both the
states of SCYM and the linearised supergravity modes expanded
around \AS. The correspondence would be no less useful as a tool
to understand SCYM theory, even if it could not then be used to
give a definition of a quantum theory of gravity away from the
limits of large $N$ and $\lth$ and small $g_{YM}^2$.

To conclude, so far as we know, present knowledge does not suffice to 
determine  which
of the three alternatives mentioned above is correct.   More work
will be required to decide the issue.

\section*{Acknowledgements}

We are very grateful to many discussions with Mike
Reisenberger during the beginning of this work.  Conversations
with Lisa Randall and Arkady Tseytlin have also been very
helpful.  Comments by Gary Horowitz, Juan Maldacena, 
K-H Rehren and Arkady Tseytlin on an
early draft of this paper have greatly helped to improve its presentation.
L.S. was supported by the NSF through grant
PHY95-14240 and gifts from the Jesse
Phillips Foundation. M.A. acknowledges support from the EU network on
``Discrete Random Geometries'', grant HPRN-CT-1999-00161.

\end{document}